# Longitudinal shaping of plasma waveguides using diffractive axicons for laser wakefield acceleration

N. Tripathi, B. Miao, A. Sloss, E. Rockafellow, J. E. Shrock, S. W. Hancock, H. M. Milchberg

[1]*Institute of Research in Electronics and Applied Physics and Dept. of Physics, University of Maryland, College Park, Maryland 20742, USA*
*milch@umd.edu

**New techniques for the optical generation of plasma waveguides—optical fibres for ultra-intense light pulses— have become vital to the advancement of multi-GeV laser wakefield acceleration. Here, we demonstrate the fabrication and characterization of a transmissive eight-level logarithmic diffractive axicon (LDA) for the generation of meter-scale plasma waveguides. These LDAs enable the formation of a Bessel-like beam with controllable start and end locations of the focal line and near-constant intensity on axis. We present measurements of the Bessel-like focal profile produced by the LDA, and of the leading end of the plasma column generated by it. One important feature is the formation of a funnel-mouthed plasma channel entrance that can act as waveguide coupler. We also compare the diffraction efficiency of our 8-level LDA to 4-level and binary versions, with measurements comparing well to theory.**

Use of RF technology for future particle accelerators faces serious challenges due to the limitation on accelerating fields imposed by material breakdown on the cavity walls. With accelerating fields up to $\sim 1000 \times$ higher produced by laser-driven plasma waves, laser wakefield acceleration (LWFA) offers a path towards the implementation of next generation accelerators and light sources [1]. Efficient LWFA of electrons to energies >1 GeV requires low plasma densities ($\lesssim 1 - 3 \times 10^{17} cm^{-3}$), and acceleration over distances many times longer than the Rayleigh range of a focused laser pulse [1,2], necessitating the use of plasma waveguides [3–11]. New techniques for optically generating meter-scale, low-density plasma waveguides [12–16] have enabled the first demonstrations of multi-GeV LWFA in optically generated plasma channels [11,17–20]. These experiments employed high-fidelity Bessel beams [21] focused by diffractive axicons (ring gratings) [14] for waveguide generation, and demonstrated that longitudinal uniformity of the waveguide can significantly impact electron bunch quality [11]. Subsequent experiments have suggested that longitudinal tailoring of the channel is desirable for further improvement in the injection and acceleration processes [18,22]. Similar to axicons, a reflective optic called an axiparabola was used to make a 10 mm long channel to guide a 20 TW beam [23], after which these waveguides were employed to accelerate electron beams to 1.1 GeV with a 50 TW beam [24].

In this paper, we present complete characterization of a customizable and efficient diffractive transmission optic [25–27] designed to improve uniformity and tailor the channel radius of plasma waveguides, the logarithmic diffractive axicon (LDA). The LDAs, as well as other diffractive axicons are fabricated at the University of Maryland using photolithography. Owing to the logarithmic variation of its phase retardance function, the focal line of an LDA has an axially decreasing spot size, giving it a trumpet/funnel shape. This trumpet-like entrance of the waveguide can potentially enhance the coupling of the guided beam into the waveguide in LWFA experiments by acting as plasma lens. This new dimension of parameter control can overcome some of the key challenges in scaling up the technology of plasma waveguides for producing multi-meter-long channels in near-future LWFA experiments.

The self-waveguiding method employed for waveguide generation in [13] relies on a Bessel beam to generate an initial plasma column via optical field ionization (OFI) [28]. The relatively cool ($\sim$10 eV) plasma column pushes radially outward, driving a single-cycle shock wave in the surrounding neutral gas. Two-color interferometry and multi-physics modeling of this process show that the expanding plasma does not itself develop a cylindrical shock, which could act as the cladding of guiding structure [13,29]. Rather, the reduced plasma density on-axis forms the low-density waveguide core, while ionization of the neutral shock by the leading wings of a high intensity pulse forms the high-density cladding, enabling self-waveguiding for the bulk of the energy in the pulse.

A class of infinite aperture $z$-propagating solutions to the Helmholtz wave equation is the Bessel beam [30,31]

$$E(r, \theta, \varphi, z) = A_0 e^{ik_z z} J_m(k_\perp r) e^{\pm im\varphi} , \qquad (1)$$

where $A_0$ is an arbitrary complex amplitude, $k_z = k\cos\gamma$ and $k_\perp = k\sin\gamma$ are longitudinal and transverse wavenumbers, $k$ is the laser wavenumber, $m \geq 0$ is an integer, $J_m$ is the $m^{th}$ order Bessel function of the first kind, $\gamma$ is the approach angle to the optical axis of the rays that form the Bessel beam, and

$|E(r,\theta,\varphi,z)|^2$ is invariant along $z$. While such infinite aperture beams are unphysical, finite aperture Bessel-like beams (which we will also refer to as Bessel beams) generated with input beams of finite diameter, say $D$, are commonly employed in a variety of applications. In such beams, the extent of axial invariance of the intensity pattern—the effective focal length— is $L = D/2\tan\gamma$. Methods for Bessel beam formation include the use of refractive and reflective axicons [3,12,32], spatial light modulators [33], and diffractive axicons (DAs) [11,14,19]. Of these approaches, DAs are best suited for the high power (> TW) and large aperture (> 5 cm) requirements for generating high intensity Bessel beams with meter-long focal lengths.

In all of these methods, Bessel beams of the form of Eq. (1) have been formed by applying a linear radial phase $\Phi(\rho) = k\rho\sin\gamma$ to the near field of the incident beam, where $\rho$ is near field radial position. This directs each annulus of the near-field, at the fixed angle $\gamma$, to a different longitudinal location on the optical axis, forming an extended focal line where the far field near the axis is well-described by the Bessel functional form of Eq. (1). For $m = 0$, the effective spot size of the Bessel beam is the radial distance to the first zero of $J_0$, $r_0 = 2.4048/k_\perp$, which is constant along the whole focal line. However, while $r_0$ is constant with $z$, the peak on-axis intensity profile, $I(z)$, is not. For an input near field intensity profile $I_{in}(\rho)$, the on-axis intensity profile is $I(z) \propto zI_{in}(\rho(z))$, where $\rho(z) = z\tan\gamma$ [4]. For a flat-top input profile, this gives $I(z) \propto z$, and for a Gaussian beam input of spot size $w_0$, $I(z) \propto z\exp(-z^2\tan^2\gamma/w_0^2)$. Since such non-uniformity of $I(z)$ can result in axial non-uniformity of optically generated plasma channels [29], techniques for controlling $I(z)$ are desirable.

Control of $I(z)$ requires control of the angle $\gamma$, or equivalently, control of the $k_\perp$ spectrum, which is a delta function for the Bessel beam of Eq. (1) [34]. In [35], Sochacki *et al.* demonstrated a method to identify the appropriate radial phase $\Phi(\rho)$ which maps $I_{in}(\rho)$ to a desired $I(z)$. To generate a uniform $I(z)$ using a flat-top input beam, the necessary radial phase imposed by the optical element, called a logarithmic or log axicon (LA), is

$$\Phi(\rho) = \frac{-k}{2a}\ln k(z_1 + a\rho^2), \quad (2)$$

where $a = (z_2 - z_1)/R^2$, $z_1$ and $z_2$ are the start and end locations of the focal line, and $R$ is the radius of the axicon.

The effect of this radial phase in the far field is a Bessel-like beam with constant $I(z)$, accompanied by a $z$-dependent spot size (for $z > z_1$),

$$r_0(z) = \frac{2.4048}{k}\left(1 + \frac{az^2}{z - z_1}\right)^{1/2}. \quad (3)$$

Logarithmic axicon-generated Bessel-like beams were first realized in [36], with a 1 cm diameter annular aperture binary phase diffractive axicon fabricated with e-beam lithography. This produced a 10 cm focal line with near-uniform intensity. However, this optic suffered from reduced diffraction efficiency of ~40% owing to the binary phase, and e-beam lithography is not suitable for scaling to large apertures. In other work [37], a 1.25 cm diameter refractive LA was fabricated using diamond turning. However, a refractive axicon is unsuitable for terawatt pulses due to phase front distortion by non-linear phase accumulation and bulk material damage. For application to generation of meter-scale plasma waveguides, diffractive axicons with thin substrates [11,14,18] are favoured.

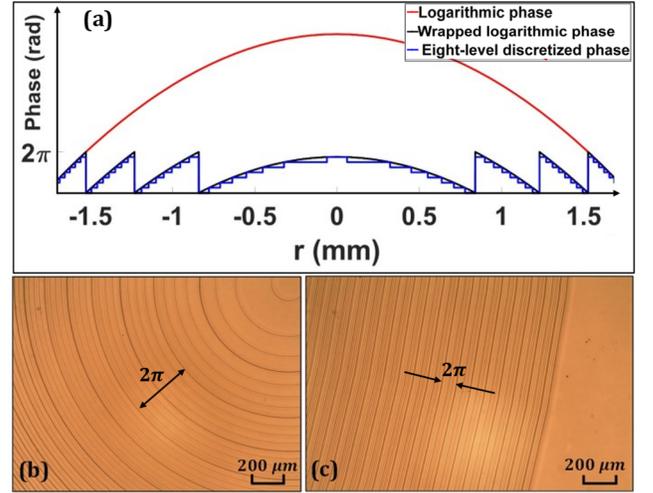

**Fig. 1. (a)** Plot showing LA phase variation along with Fresnel LA (wrapped phase) and eight phase level (or 8-step) LDA. Microscope images of ring grating phase structure **(b)** near the center of the fabricated log axicon, and **(c)** near the axicon edge.

The design process for diffractive axicons suitable for meter-scale, guided LWFA is illustrated in Fig. 1(a). The phase profile determined from Eq. (2) (red curve) is wrapped onto the $[0, 2\pi]$ interval (black curve), analogous to Fresnel lens. For lithographic fabrication, this phase profile is discretized into $N$ steps (shown for $N = 8$ in the blue curve) per $2\pi$ interval. Each step of this staircase pattern provides a phase difference of $\pi/4$, corresponding to a step height of 2206 Å, optimized for a central wavelength of 800 nm. This multi-surface, variable period ring grating is fabricated using photolithography in the FabLab at the University of Maryland. Use of photolithography allows the fabrication time and cost to stay manageable when scaling to larger diameter DAs and LDAs.

For an experimental demonstration of the flat axial profile, we fabricated a LDA with an aperture diameter of 2 cm with $z_1 = 50$ cm and $z_2 = 100$ cm (focal length 50 cm). The radial phase of Eq. (2) leads to progressively shorter radial periods of $2\pi$ phase variation toward the outside of the axicon, as seen in the microscope images of the fabricated log axicon in Figs. 1(b) and 1(c).

We simulated the on-axis $I(z)$ profile produced with our LA with our unidirectional pulse propagation code YAPPE [29], using an input beam with the experimental parameters (central wavelength $\lambda_0 = 800$ nm, 35 fs pulsewidth, 30 nm bandwidth) and assuming a flat-top intensity profile, The result is plotted as the red curve in Fig. 2(a). In the experiment, the near-field mode of the input beam was expanded and collimated to overfill the 2 cm aperture of the optic, forming a near flat-top profile. An axial image scan of the focal region was taken with a 10 × microscope objective and CCD camera mounted on a 75 $cm$ long motorized stage. Five images were averaged at each axial position. Camera calibration was performed using a fused silica conical refractive axicon with known ray approach angle $\gamma = 4°$. The concatenated image is shown in Fig. 2(b).

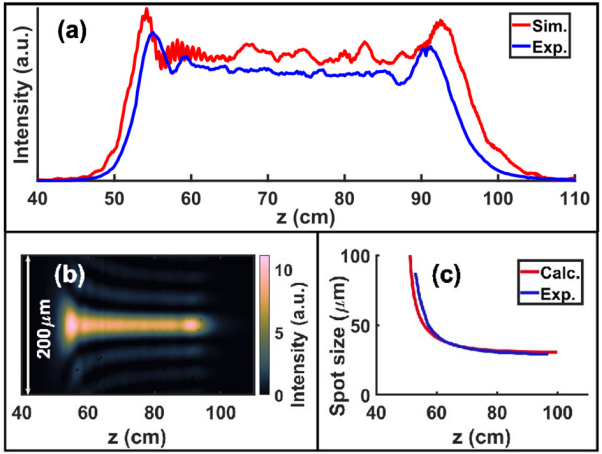

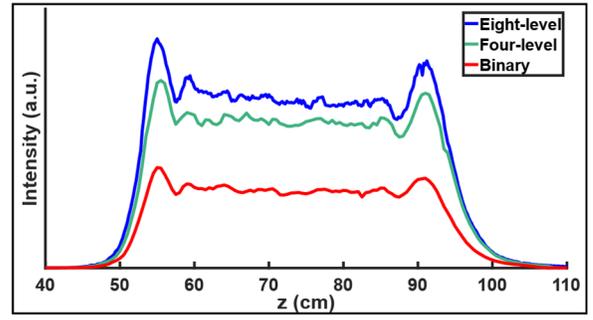

**Fig. 2.** (a) Measurement of axial intensity from an 8-level LDA of 2 cm aperture and 50 cm focal line length along with the simulated intensity with the same axicon parameters. (b) Horizontal slice of Bessel beam intensity pattern vs. position along focal line. (c) Calculated and experimentally recorded spot size variation of the Bessel beam pattern.

**Fig. 3.** Comparison between axial intensity due to binary, four-level, and eight-level LA, each with the diameter of 2 cm and focal length of 50 cm starting with first focus at 50 cm. The measured ratio of mean intensities $I_8/I_4/I_2$ is 1/0.87/0.46 while Eq. (4) predicts 1/0.854/0.427.

The peak intensity variation $I(z)$, taken as the central lineout of Fig. 2(b), is plotted as the blue curve in Fig. 2(a). It is in reasonable agreement with the simulation, showing intensity peaks near $z_1 = 50 \, cm$ and $z_2 = 100 \, cm$ and a flat intensity profile in between. The intensity peaks at the beginning and end of the focal line are due to diffraction of the incident beam from the inner and outer edges of the ring grating pattern. In addition, oscillations between the end peaks are evident, with less resolution in the measurement than in the simulation. These oscillations are caused by interference between the rays of the beam passing through the centre of the LDA and the axicon-diffracted rays with varying $k_z = k \cos\gamma(z)$. The leading peak can be suppressed using an annular aperture LA [36,38] to block the central part of the input beam. This is accomplished in our LWFA experiments [11,19], where the central part of the channel-forming beam is lost when reflected using a mirror with a central drilled hole to allow the accelerator drive beam to couple into the waveguide. The trailing peak can be reduced by using soft-edged apodizers on the near field beam [39] or a Gaussian or super-Gaussian input.

The widening of the Bessel beam's central maximum toward $z_1 = 50 \, cm$, as seen in Fig. 2(b), is predicted by Eq. (3), which is plotted as the red curve in Fig. 2(c). Overlaid as the blue curve are the measured values of $r_0(z)$, obtained by fitting the square of the zeroth order Bessel function to the measured focal profiles. Agreement with theory is excellent.

The primary drawback of diffractive optics compared to reflective/refractive optics is possible reduced efficiency. However, increasing the number of phase levels or steps increases the efficiency. The theoretical first order diffraction efficiency for an $N$ step grating is [25]

$$\eta(N) = \left[\frac{\sin(\pi/N)}{\pi/N}\right]^2, \quad (4)$$

which predicts $\eta(N = 8) = 95\%$. The measured efficiency of 87% is slightly lower, likely due to the $\sim 1 \, \mu m$ alignment uncertainty of the lithographic masks used for etching the many discrete steps highlighted in Fig. 1. To confirm higher efficiency in the 8-level optic employed for the measurements in Fig. 2, we fabricated additional LDAs with 4- and 2-step (binary) profiles. Figure 3 compares the measured on-axis intensity profiles $I(z)$ from these optics. If the mean intensity $I_8$ of the 8-level curve in Fig. 3 is taken as unity, then the measured ratio $I_8/I_4/I_2$ is 1/0.87/0.46 while Eq. (4) predicts 1/0.854/0.427, in good agreement. These efficiency ratios are comparable to those of our 4- and 8-level *linear* diffractive axicons fabricated using the same method and employed in [11,14,18].

The combination of constant intensity and the wider Bessel beam profile at the leading end of the focal line (Fig. 2(b)) results in increased energy per unit length deposited there in optical field ionization (OFI) and plasma heating. This increases the local rate of hydrodynamic expansion [40], also observed more recently by measuring faster expansion from wider initial OFI plasma columns [13,29]. This effect leads to a wider plasma channel near the leading end, forming an end-funnel for enhanced plasma waveguide entrance coupling, previously demonstrated with a separate laser pulse heating the end of a plasma waveguide [41,42]. In principle, the funnel acts as a mode converter (akin to a tapered optical fiber), improving tolerance to misalignment.

To demonstrate formation of a funnel shaped plasma, we directed a 150 mJ, 50 fs, λ=800nm, 2.4 cm diameter input pulse through an LDA ($R = 3.4 \, cm$, $z_1 = 20 \, cm$, $z_2 = 80 \, cm$) and focused in a 50 Torr H$_2$ backfill. A 50 fs, λ=800nm probe pulse was passed transversely through the plasma column and then through a folded wavefront interferometer. See Fig. 4(a). Interferometer measurements were recorded by stepping through the first 100 mm of the focal line. At each position 50 shots were taken, with column averaging sufficiently minimizing noise for measurement of the ~0.1 mrad phase shifts imparted by the low-density plasmas. Abel inversion of the extracted phase enables reconstruction of the transverse electron density profiles [13,40]. The leading portion of the plasma profile is plotted in Fig. 4(b), clearly showing the wide funnel, and confirming the predicted effect of the intensity profile of Fig. 2(b). The longitudinal oscillations in the density originate from uncorrected radial intensity modulations in the incident beam, which resulted in a varying level of OFI in the Bessel beam focus.

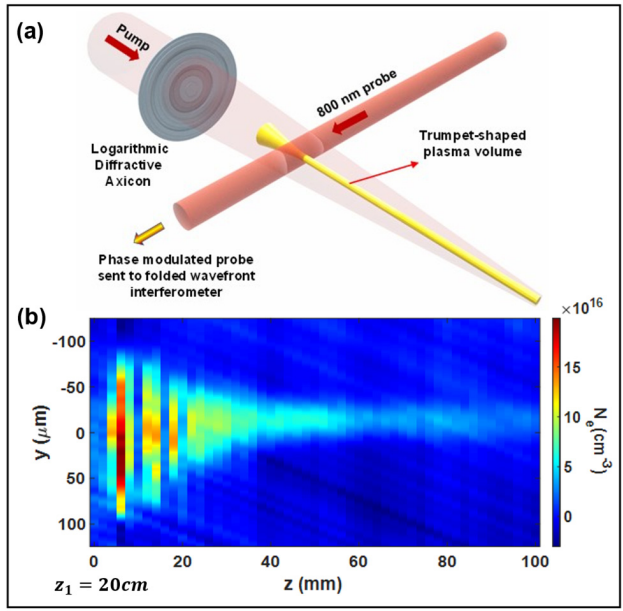

**Fig. 4.** (a) Experimental setup; (b) Leading portion of the electron density profile generated using a logarithmic diffractive axicon. The horizontal scale is referenced to $z_1 = 20\ cm$.

To summarize, we have presented an approach for longitudinal shaping of plasma waveguides by tailoring the properties of Bessel-like beams. Diffractive transmission axicons are exceptionally well-suited to this application. We presented theory, simulations, and measurements for a case of particular interest, the logarithmic diffractive axicon, designed to produce a uniform on-axis intensity distribution. Such a distribution is necessarily accompanied by an axial position-dependent Bessel beam central spot size, which we employ to demonstrate generation of a plasma funnel at the entrance of a plasma channel. We expect such a plasma funnel to lead to enhanced laser coupling and guiding in laser wakefield acceleration experiments. For laser accelerators to 10 GeV and beyond, which require plasma waveguides of several meters in length, we expect that customizable diffractive axicons will become indispensable elements of primary importance.

**Funding.** This work was supported by the U.S. Department of Energy (DE-SC0015516), and the Defense Advanced Research Projects Agency (DARPA) under the Muons for Science and Security Program.

**Acknowledgments**. The authors acknowledge valuable assistance from Nam Kim and Mark Lecates at the University of Maryland FabLab.

**Disclosures.** The authors declare no conflicts of interest.

**Data availability.** Data underlying the results may be obtained from the authors upon reasonable request.